\documentclass[twocolumn,showpacs,preprintnumbers,amsmath,amssymb,english,aps]{revtex4}
\usepackage[T1]{fontenc}
\usepackage[latin1]{inputenc}
\usepackage{amsmath}
\usepackage{graphicx}
\usepackage{amssymb}

\makeatletter

\providecommand{\LyX}{L\kern-.1667em\lower.25em\hbox{Y}\kern-.125ems\@}

\usepackage{graphicx}
\usepackage{graphicx}
\usepackage{graphicx}

\usepackage{graphicx}

\usepackage{graphicx,amsfonts}
\usepackage{dcolumn}
\usepackage{bm}
\usepackage{babel}
\makeatother

\newcommand{\beq}{\begin{equation}} \newcommand{\eeq}{\end{equation}}
\newcommand{\beqa}{\begin{eqnarray}}
\newcommand{\eeqa}{\end{eqnarray}}
\newcommand{\beqam}{\begin{eqnarray*}}
\newcommand{\eeqam}{\end{eqnarray*}}

\newcommand{\ket}[1]{|#1\rangle}

\begin{document}

\title{An algorithm to perform POVMs through Neumark theorem:
application to the discrimination of non-orthogonal pure 
quantum states}

\author{Wilson R. M. Rabelo}
\email{wilson@unifap.br}
\author{Alexandre G. Rodrigues}
\author{Reinaldo O. Vianna}
\email{reinaldo@fisica.ufmg.br}
\affiliation{ Grupo Informa\c c\~ao Qu\^antica - Departamento de F\'{\i}sica - ICEX -CP 702 - Universidade
Federal de Minas Gerais - 30123-970 - Belo Horizonte - MG -
Brazil}
\date{\today}

\begin{abstract}
We consider a  protocol to perform the optimal quantum state discrimination of
 $N$ linearly independent non-orthogonal pure quantum states and present a
computational code.
Through the extension of  the original  Hilbert space,  
it is possible to perform  an unitary operation yielding  a final configuration,
which gives the best discrimination  without ambiguity by means of von Neumann measurements.
Our  goal is to introduce a detailed general mathematical procedure to realize this task 
by means of  semidefinite programming and norm minimization.
 The former is used  to fix which is the best detection 
probability amplitude for each state of the ensemble. The latter determines the
 matrix which leads the states to the final configuration. In a final step,
 we decompose the unitary transformation  in a  sequence of two-level
 rotation matrices. 
\end{abstract}

\maketitle

\section{Introduction}

Quantum  state discrimination is a crucial problem in quantum information theory, 
and it  is well known the impossibility of doing this  perfectly for  non-orthogonal states \cite{Peres}.
 Nevertheless, if a non zero probability of inconclusive results is allowed for, 
it is possible to never mistake a state for another, by means of an appropriate Positive
Operator Valued Measure  (POVM). 
This strategy is known as  unambiguous state discrimination (USD), 
and the best procedure of this kind is that which minimizes the probability of inconclusive results.
The  discrimination of two equally probable non-orthogonal pure states was firstly considered
 by Ivanovich \cite{Iva}, Dieks \cite{Dieks}  and Peres \cite{Peres2}. 
The case of two states with unequal prior probabilities was treated by Jaeger and Shimony \cite{Jaeger}.
Chefles \cite{Chefles} showed that USD of $N$ pure quantum states is possible  if and
 only if they are linearly independent. 

With the aid of  Semidefinite Programming (SDP),  Eldar  \cite{YO-2002} obtained 
the set of necessary and sufficient conditions the USD measurement operators satisfy.
She also showed that for a given set of pure states, there always is an ensemble
for which the optimal USD detects each state with equal probability (i.e, an
Equal Probability Measurement - EPM).

In references \cite{Hillery1} and \cite{Saavedra}, the USD problem is
approached via the Neumark theorem \cite{Peres}, i.e, the realization of
a POVM by means of projective measurements in an extended Hilbert space.
In \cite{Hillery1},
it is done in the context of linear quantum optics , whereas in \cite{Saavedra} 
an ions trap architecture is considered. 

Our main goal is to derive the transformation which maps $N$ 
non-orthogonal pure states in a set of states that can be
discriminated by usual projective measurements in an extended
Hilbert space. It is equivalent to a generalized measurement
in the original space, and is the content of Neumark's theorem.
We then present a simple MATLAB code that takes as input the
ensemble of non-orthogonal states and outputs the best set of discriminable
states and the pertinent transformations. We believe this  computational
code can be pedagogically useful to anyone getting started with quantum
information and semidefinite programming.

The paper is organized as follows. In section I we discuss generalized
measurements and semidefinite programming. In section II we present
a protocol to discriminate pure non-orthogonal states and, in section
IV, we exercise with a numerical example in the 
context of quantum key distribution. We conclude in section V.
In the appendix, we furnish our MATLAB code.

\section{POVMs AND DISCRIMINATION WITH SDP}

The operation commonly known as  
 positive operator-valued measure  can be performed
 with a set of quantum detection operators $\Pi_k$, where
 the probability $p_k$ of obtaining the state labeled by 
the index $k$ is given by
\begin{equation}
p_k=Tr(\Pi_k\rho),
\end{equation}
where $\rho$ is the density operator of the system. As the
probabilities are obviously  non-negative reals  and sum
 to one,
 all the quantum detection operators are semidefinite
 operators and form a resolution of the identity,
\begin{equation}
\sum_k\Pi_k=I.
\label{condition}
\end{equation}
The problem of USD for $N$ pure non-orthogonal states 
can be stated as follows. We assume that a quantum system was prepared in one of the $N$ pure states $\{\left|Q_i\right\rangle\}$, $i=1,...,N$. Each state $\left|Q_i\right\rangle$ is in a $N$-dimensional Hilbert space. 
 In order to identify a state or to return an inconclusive
 result, the measurement operators must obey
 $\left\langle Q_i\right|\Pi_k\left|Q_i\right\rangle=p_i\delta_{ik}$ with $0\leq{p_i}\leq{1}$. 
Therefore, to  satisfy  condition (\ref{condition}), we have $\sum_{i=0}^{N}\Pi_i=I$ and the inconclusive result is given by $\Pi_0=I-\sum_{i=1}^{N}\Pi_i$.

In \cite{YO-2002}, Eldar showed that the optimal unambiguous discrimination can be formulated  as a SDP problem. The measurement operators are expressed  in the form
\begin{equation}
\Pi_i=p_i|\tilde{Q}_i\rangle\langle\tilde{Q}_i|,
\label{reciprocal}
\end{equation}
where each state $|\tilde{Q}_i\rangle$ is in a $N$-dimensional
 Hilbert space, and the $\Pi_i$ are not orthogonal projectors 
in this space.  The vectors $|\tilde{Q}_i\rangle$ are the
 \textit{reciprocal states} associated with  $|Q_i\rangle$, 
such that
\beqa
\langle\tilde{Q}_i|Q_k\rangle=\zeta_{ik}\delta_{ik}\;\;,\;\;1\leq i,k\leq N
\eeqa
where the factor $\zeta_{ik}$ indicates the scalar product is not 
normalized. Given the matrix $\Psi$, whose columns are the
vectors  $|Q_i\rangle$, the $|\tilde{Q}_i\rangle$ are the
columns of the matrix $\tilde{\Psi}$, namely,

\beqa
\tilde{\Psi}=\Psi(\Psi^*\Psi)^{-1}.
\label{matrix-psi}
\eeqa

Given an ensemble $\rho$, where each state $|Q_i\rangle$ is
prepared with probability $\mu_i$, the total probability
of a successful detection is
\beqa
P_D=\sum_{i=1}^N\mu_i\langle Q_i|\Pi_i|Q_i\rangle=\sum_{i=1}^N\mu_ip_i.
\label{max_Pd}
\eeqa
The problem of optimal USD is then  to find measurement 
operators $\Pi_i$, or equivalently, the probabilities $p_i$ ,
which maximize $P_D$, subject to the constraint 
(\ref{condition}), which can be recast as 
\begin{equation}  
I-\sum_{i=1}^{N}p_i|\tilde{Q}_i\rangle\langle\tilde{Q}_i|\geq 0 .
\label{constraint}
\end{equation}
A  SDP problem is to find  $x\in R^m$, which minimizes
the linear function $c^{T}x$ , subject to the  matrix 
inequality $F(x)=F_0+\sum_{i=1}^{m}x_iF_i\geq 0$,
 where the problem data are the vector $c\in R^m$ and
 the $m+1$ complex Hermitian matrices $F_i$ \cite{Boyd}.
 It is known as the primal formulation of SDP.

Equations (\ref{max_Pd}) and (\ref{constraint}) can be 
recast as a SDP problem, namely:  
\beqa
\mathrm{min}_{p\in R^N} \{-\mu^T{p}\},
\label{primal}
\eeqa
subject to $N+1$ constraints,
\beqa
I-\sum_{i=1}^N p_i|\tilde{Q}_i\rangle\langle\tilde{Q}_i| \geq 0,\nonumber\\ 
p_i\geq0,\;\;1\leq i\leq N.
\label{constraints-N}
\eeqa

\section{Protocol for  optimal discrimination of pure states via rotations and von Neumann measurements}

We start out by rewriting the $N$ entry states to be discriminated
in a ladder form in the orthonormal basis 
 $\{ \left|i\right\rangle, i=1,\hdots,N\}$,
\beqa
\left|Q_1\right\rangle&=&\left|1\right\rangle, \nonumber\\
\left|Q_2\right\rangle&=&c_{12}\left|1\right\rangle+ c_{22}\left|2\right\rangle, \nonumber\\
\left|Q_3\right\rangle&=&c_{13}\left|1\right\rangle+ c_{23}\left|2\right\rangle+ c_{33}\left|3\right\rangle \nonumber\\
\vdots\phantom{cc}&&\nonumber\\
\left|Q_N\right\rangle&=&c_{1N}\left|1\right\rangle+ \hdots+ c_{N\!N}\left|N\right\rangle~~,
\label{Q-inicial}
\eeqa
where $\{\left|Q_i\right\rangle\}\in{\cal{H}}_N$ ($N$-dimensional Hilbert space).
This can always be done by means of a unitary transformation
$U_0$.
Then, to apply Neumark's theorem, we extend the original 
Hilbert space, e.g. through addition of ancillas,
to $2N-1$ dimensions and map the original states to the
{\em final configuration},

\beqa
\label{Qf0}
\left|Q_{1f}\!\right\rangle&=&g_1\left|1\right\rangle\! + \!g_{N\!\!+\!1}\!\left|N\!\!+\!\!1\right\rangle\!+\!\!\hdots\!+\!g_{2N\!\!-\!1}\!\left|2N\!\!-\!\!1\right\rangle\nonumber\\
\left|Q_{2f}\!\right\rangle&=&g_2\left|2\right\rangle\!+\! g_{2N}\!\left|N\!\!+\!\!1\right\rangle\!\!+\!\! \hdots\!+\!g_{3N\!-\!2}\!\left|2N\!\!-\!1\right\rangle\nonumber\\
\left|Q_{3f}\!\right\rangle&=&g_3\!\left|3\right\rangle\!+\! g_{3N\!-\!1}\!\left|N\!\!+\!\!1\right\rangle\!\!+\!\!\hdots\!\!+\!\!g_{4\!N\!-\!4}\!\!\left|2N\!\!-\!2\right\rangle\nonumber\\
\vdots\phantom{cc} & & \phantom{cccccccccc}\vdots\nonumber\\
\ket{Q_{if}}&=&g_i\ket{i}+g_{[\frac{i}{2}(2N+3-i)-1]}\ket{N+1}+
 \nonumber\\
& &\hdots+ g_{[N(i+1)+\frac{i}{2}(1-i)-1]}\ket{2N+1-i} \nonumber\\
\vdots\phantom{cc} & & \phantom{cccccccccc}\vdots\nonumber\\
\left|Q_{N\!f}\!\right\rangle&=&g_N\!\left|N\!\right\rangle+  g_{[\frac{1}{2}N(N\!\!+\!3)\!-\!1]}\!\left|N\!+\!1\right\rangle.
\eeqa
Now, a projective measurement in the orthonormal basis
$\{ \left|i\right\rangle, i=1,\hdots,2N-1\}$ yields an
unambiguous discrimination. The state labeled by $i$ is
identified with probability $g_i^2$, when the measurement
collapses to $\left|i\right\rangle$ for $1\leq i \leq N$.
For other values of $i$ the result is inconclusive.
Therefore, the first $N$ $ g_i$ are chosen to produce
the best possible discrimination, i.e, $g_i=\sqrt{p_i}$ for
the $p_i$ defined in Eq.3, which are determined by SDP.
The other coefficients $\{ g_i, 
i=N+1$ to $\frac{N}{2}(N+3)-1\}$ are fixed in order to preserve normalization
and the scalar products among the original states.

Once the final configuration is known, we want to determine
the unitary transformation $U_1$ which maps the original 
states to it. Norm minimization \cite{Horn} implies,
\beqa
\label{norm-problem}
\!\!\!\!\!\!\!\!\!\!\!\!\sum_{i=1}^N\left\|\left|Q_{if}\!\right\rangle-\!U_1\!\left|Q_i\right\rangle\right\|^2=2N-2Re \left[Tr(A'U_1^{\dag})\right]=0,
\eeqa
where
\beqa
\label{A_linha}
A^\prime=\sum_{i=1}^N\left|Q_{if}\rangle\langle Q_i\right|.
\eeqa
$A'$ is a singular  $(2N\!\!-\!\!1)\!\times\!(2N\!\!-\!\!1)$ complex matrix.
The singular value decomposition (SVD) of $A'$ is 
$A'=V\Sigma W^{\dag}$, with $V$ and
 $W$ $(2N\!\!-\!\!1)\!\times\!(2N\!\!-\!\!1)$ unitary matrices,
 and  $\Sigma$  a diagonal $(2N\!-\!1)\!\times\!(2N\!-\!1)$
 matrix,  then $U_1$ is given by
\beqa
U_1=VW^\dagger.
\label{U_best}
\eeqa

The unitary transformation $U=U_1U_0$  can now be decomposed \cite{Reck} 
in a sequence of rotations ($R_{kl}$) in the 
hyperplanes ($kl$), as
\beqa
\label{produto}
U=\prod_{k=1}^{N-1}\prod_{l=k+1}^{N}R_{kl}^{\dagger}.
\eeqa
The $R_{kl}$ are two-level matrices, with the four 
non trivial entries, 
\[ [R_{kl}]_{kk}=\frac{[U]_{kk}^*}{\sqrt{|[U]_{kk}|^2
+ |[U]_{lk}|^2}},\]
\[ [R_{kl}]_{ll}=-[R_{kl}]_{kk}^*, \]
\[ [R_{kl}]_{kl}=\frac{[U]_{lk}}{\sqrt{|[U]_{kk}|^2
+ |[U]_{lk}|^2}},\]
\[ [R_{kl}]_{lk}=[R_{kl}]_{kl}^*.\]
The trivial entries are
\[ [R_{kl}]_{mm}=1, m\neq (k,l),\]
\[ [R_{kl}]_{mn}=0, (m,n)\neq (k,l).\]

We can also express the rotations $R_{kl}^{\dagger}$
 in terms of  Pauli  matrices,  $R_z^k(\theta)=\exp(-i\theta\sigma_z/2)$ (the super-index $k$ 
indicates the rotation is over the ket
 $ \left|k\right\rangle$), $R_y^l(\theta)=\exp(-i\theta\sigma_y/2)$. Hence, 
\beqa
R_{kl}^{\dagger}(\alpha,\beta,\gamma,\delta)=e^{i\alpha}R_z^{k}(\beta)R_y^{l}(\gamma)R_z^{k}(\delta).
\label{Rgeral}
\eeqa

Therefore the unitary transformation $U=U_1U_0$, followed by
a projective measurement in the basis 
$\{\ket{i}, i=1,...,2N-1\}$, discriminates unambiguously any $N$ pure non-orthogonal states.
Summarizing, we have the following algorithm:

\begin{itemize}
\item Rewrite the entry states in a ladder form $(U_0)$.
\item Fix the  {\em conclusive amplitudes} ($g_1$ to
$g_N$) using SDP.
\item  Fix the {\em inconclusive amplitudes} ($g_{N+1}$ 
to $g_{\frac{N}{2}(N+3)-1}$) such that normalization and
scalar products among the states be preserved.
\item Build the unitary transformation ($U_1$) that maps
the entry states in the ladder form to the {\em final 
discriminable configuration}.
\item Decompose $U=U_1U_0$ as a product of one-qudit rotations.
\end{itemize}

\section{Example}

As an example of application of our code, we consider
how a spy (Eve) could use USD to eavesdrop 
\cite{Dusek, Enk} two parties (Alice and Bob) 
establishing a BB84 \cite{Bennet} cryptographic key.

Alice and Bob are communicating by means of low
intensity laser pulses in a network of lossy optical
fibers.
The signals sent by Alice are coherent states with
a certain mean number of photons ($\mu$). 
The polarization of the pulses are randomly chosen
with equal probability among four possibilities, 
namely, diagonal to right or left ($\ket{d+}$ , 
$\ket{d-}$), and circularly polarized to
right or left ($\ket{c-}$ ,
     $\ket{c+}$).
Bob directs the pulses he receives to detectors,
which do not distinguish photon numbers, preceded
by polarization analyzers which can be either for
linear or circular polarization, also chosen 
randomly with equal probability.
After a certain number of pulses, Alice publicly
announces the sequence of polarization basis
she used. Bob then checks which pulses he 
detected using the compatible polarization 
analyzer. After discounting dark counts and losses
in the fibers, the matching detections allow Alice
and Bob to establish a cryptographic key, i.e., a 
long sequence of zeroes and ones (say $\ket{d+}$, 
$\ket{c+}$ for one and $\ket{d-}$, $\ket{c-}$ for
zero). This is the very well known BB84 protocol.

Eve could try to obtain this key as follows. 
She probes the network and measures the mean
number of photons in the pulses, which can
be done without disturbing the polarization. 
Now she knows that most of the time Bob receives
states of the type $\ket{polarization}^{\otimes \mu}$.
Suppose $\mu$ is 3. Eve then prepares an USD scheme
for the states $\ket{d+}\ket{d+}\ket{d+}$, 
$\ket{d-}\ket{d-}\ket{d-}$, $\ket{c+}\ket{c+}\ket{c+}$, 
$\ket{c-}\ket{c-}\ket{c-}$.

Running our code, all the pertinent parameters to prepare
the USD are yielded. In particular, we learn that Eve 
correctly identifies the polarization state with a 
probability of $50\%$, if the state has three photons.
When she succeeds, she prepares a state with the correct
polarization and sends it to Bob, who will never know
it came from Eve. If she fails, she does nothing at all, 
and Bob could think it is a dark count or a loss in
the network.
\begin{widetext}
\beqa
\!\!\!\!\!\!\!\!\!\!\!\!\!\!\!\!\!\!\!\!\!\!\!\!\!\!\!\!\!\!U_0 \!\!=\!\!\left[ \begin{array}{crrrrrrr}
0.3536& 0.3536 & 0.3536  &   0.3536 &   0.3536 &   0.3536 &   0.3536 &  0.3536 \\
0.3536& -0.3536&-0.3536  &   0.3536 &  -0.3536 &   0.3536 &   0.3536 & -0.3536  \\
0.6124&-0.2041i&-0.2041i &  -0.2041 &  -0.2041i&  -0.2041 &  -0.2041 &  0.6124i \\
0.6124& 0.2041i& 0.2041i &  -0.2041 &   0.2041i&  -0.2041 &  -0.2041 & -0.6124i \\
 0    & -0.8165& 0.4082  &     0    &   0.4082 &      0   &   0.0000 &     0    \\
 0    &  0     &-0.1736  &  -0.5414 &   0.1736 &   0.7707 &  -0.2293 &     0   \\
 0    &  0     &-0.1736  &  -0.5414 &   0.1736 &  -0.2293 &   0.7707 &     0    \\
 0    &  0     & 0.6631  &  -0.2836 &  -0.6631 &   0.1418 &   0.1418 &     0  
\end{array} \right] \nonumber.
\eeqa  
\end{widetext}

Let's consider  the four  quantum states
\beqa
&&|Q_1^i\rangle=|d+\rangle  |d+\rangle  |d+\rangle \\
&&|Q_2^i\rangle=|d-\rangle  |d-\rangle  |d-\rangle\\
&&|Q_3^i\rangle=|c+\rangle  |c+\rangle  |c+\rangle \\
&&|Q_4^i\rangle=|c-\rangle  |c-\rangle  |c-\rangle ~~~,
\eeqa
where 
\beqa
&&|d+\rangle=(|0\rangle+|1\rangle)/\sqrt{2} \nonumber\\
&&|d-\rangle=(|0\rangle-|1\rangle)/\sqrt{2} \nonumber\\
&&|c+\rangle=(|0\rangle+i|1\rangle)/\sqrt{2} \nonumber\\
&&|c-\rangle=(|0\rangle-i|1\rangle)/\sqrt{2} \nonumber~~~.
\eeqa
Rewrite the entry states in a ladder form
\beqa
\!\!\!\!|Q_1\rangle\!\!\! &=&\!\!\!|1\rangle                                  \nonumber \\
\!\!\!\!|Q_2\rangle\!\!\! &=&\!\!\!|2\rangle           \nonumber \\
\!\!\!\!|Q_3\rangle\!\!\! &=&\!\!\!-(0.25-0.25i)|1\rangle +              \nonumber \\
                 \!\!\!\! &&\!\!\!-(0.25+0.25i) |2\rangle +0.8660|3\rangle \nonumber\\  
\!\!\!\!|Q_4\rangle\!\!\! &=&\!\!\!-(0.25+0.25i)|1\rangle +              \nonumber\\
                 \!\!\!\! &&\!\!\!-(0.25-0.25i) |2\rangle +0.8660|4\rangle ,\nonumber\\
\eeqa
where the unitary matrix $U_0$ for the transformation is given above.
 
The best conclusive probability amplitude given by the SDP technique is given by $g_1$$=$$g_2$$=$$g_3$$=$$g_4$$=$$0.7071$, where the weight of each state in the ensemble is $\mu_1$$=$$\mu_2$$=$$\mu_3$$=$$\mu_4$$=$$0.25$. The inconclusive amplitudes ($g_5$ to $g_{13}$) are fixed such that normalization and
scalar products among the states be preserved. For this case, $g_5=-0.3536+0.3536i$, $g_6=-0.3536-0.3536i$, $g_7=0$, $g_8=-0.3536-0.3536i$, $g_9=-0.3536+0.3536i$, $g_{10}=0$, $g_{11}=0$, 
$g_{12}=0.7071$ and $g_{13}=0.7071$.

The  final  discriminable  configuration in the extended Hilbert space is
\begin{eqnarray}
|Q_{1f}\rangle\!\!&=&0.7071|1\rangle-(0.3536-0.3536i)|5\rangle  \nonumber \\
                    &&-(0.3536+0.3536i)|6\rangle  \nonumber \\
|Q_{2f}\rangle\!\!&=&0.7071|2\rangle-(0.3536+0.3536i)|5\rangle \nonumber \\
                    &&-(0.3536-0.3536i)|6\rangle  \nonumber \\
|Q_{3f}\rangle\!\!&=&0.7071|3\rangle+ 0.7071|6\rangle    \nonumber \\
|Q_{4f}\rangle\!\!&=&0.7071|4\rangle+ 0.7071|5\rangle ~~. 
\label{4-finais-exe}
\end{eqnarray}

Now we build the unitary transformation ($U_1$) that maps the entry states in the ladder form to the final 
discriminable configuration. Following the procedure described by equations (\ref{A_linha}) and 
(\ref{U_best}), we obtain:
\begin{widetext}
\beqa
\!\!\!\!\!\!\!\!\!\!\!\!\!\!\!\!\!\!\!\!\!\!\!\!\!\!U_1\!\!=\!\!\left[ \begin{array}{rrrrrrrr}
0.7071          &    0          &0.2041\!\!-\!\!0.2041i&  0.2041\!\!+\!\!0.2041i&   0.1608\!\!+\!\!0.2396i& 0.3757\!\!-\!\!0.3300i&0&0 \\
0               &    0.7071     &0.2041\!\!+\!\!0.2041i&  0.2041\!\!-\!\!0.2041i&   0.4884\!\!-\!\!0.1074i& 0.0979\!\!+\!\!0.2715i&0&0 \\
0               &    0          &0.8165        &  0               &  -0.1511\!\!+\!\!0.0977i&-0.5375\!\!-\!\!0.1097i&0&0 \\
0               &    0          &  0           &  0.8165          &  -0.4981\!\!-\!\!0.2299i& 0.0639\!\!+\!\!0.1681i&0&0 \\
-0.3536\!\!+\!\!0.3536i &-0.3536\!\!-\!\!0.3536i&  0           &  0.4082          &   0.4981\!\!+\!\!0.2299i& 0.0639\!\!-\!\!0.1681i&0&0  \\
-0.3536\!\!-\!\!0.3536i &-0.3536\!\!+\!\!0.3536i&0.4082        &  0               &   0.1511\!\!-\!\!0.0977i& 0.5375\!\!+\!\!0.1097i&0&0  \\
0               &    0          &   0          &  0               &         0         &       0         &1&0 \\
0               &    0          &   0          &  0               &         0         &       0         &0&1  
\end{array} \right].\nonumber
\eeqa
\end{widetext}

As the last step, we wish to decompose the resultant matrix $U=U_1U_0$ as a product of one-qudit rotations. Therefore, we have

\begin{eqnarray}
\!\!\!\!\!\!U\!\!=\!\!R_{1,2}^{\dagger}R_{1,3}^{\dagger} \ldots R_{7,8}^{\dagger}
\label{rotacoes-geral2}
\end{eqnarray}

where $R_{1,5}^{\dagger}\!\!=\!\!R_{1,6}^{\dagger}\!\!=\!\!R_{1,7}^{\dagger}\!\!=\!\!R_{1,8}^{\dagger}\!\!=\!\!R_{2,7}^{\dagger}\!\!=\!\!R_{2,8}^{\dagger}\!\!=\!\!I$,
\begin{eqnarray}
\!\!\!\!R_{1,2}^{\dagger}\!\!=\!\!\left[ \begin{array}{ccccc}
    0.7071  &       0.7071 &  0     &    \cdots      &    0     \\
    0.7071  &       -0.7071&  0     &    \cdots      &    0     \\
        0   &         0    &  1     &    \cdots      &    0     \\
    \vdots  &     \vdots   & \vdots &    \ddots &   \vdots \\
        0   &          0   &  0     &    \cdots      &    1        
 \end{array}\right], \nonumber
\label{R12}
\end{eqnarray}
\begin{eqnarray}
\!\!\!\!R_{1,3}^{\dagger}\!\!=\!\!\left[ \begin{array}{ccccc}
    0.8165   &        0   &  0.5774       &    \cdots      &    0     \\
      0      &          1 &  0            &    \cdots      &    0     \\
    0.5774   &     0      &  -0.8165      &    \cdots      &    0     \\
    \vdots   &     \vdots & \vdots        &    \ddots &   \vdots \\
        0    &          0 &  0            &    \cdots      &    1  
 \end{array}\right], \nonumber
\label{R13}
\end{eqnarray}

\begin{eqnarray}
\!\!\!\!R_{1,4}^{\dagger}\!\!=\!\!\left[ \begin{array}{cccccc}
    0.8660    &  0    & 0    &   0.5     &\ldots   & 0     \\
      0       &  1    &  0   &    0      &\ldots   & 0     \\
      0       &  0    &  1   &    0      &\ldots   & 0     \\
    0.5       &  0    &  0   &   -0.8660 &\ldots   & 0     \\
  \vdots      &\vdots &\vdots&   \vdots  &\ddots   & \vdots    \\
      0       &  0    &  0   &    0      & \ldots  & 1       
\end{array}\right], \nonumber
\label{R14}
\end{eqnarray}

\begin{eqnarray}
\!\!\!\!\!\!\!\!\!\!\!\!\!\!\!\!R_{2,3}^{\dagger}\!\!=\!\!\left[ \begin{array}{ccccc}
        1&    0                    &    0                    & \ldots \\     
        0& 0.7224\!\!-\!\!0.3407i  & -0.5392\!\!-\!\!0.2671i & \ldots  \\         
        0&  -0.5392\!\!+\!\!0.2671i& -0.7224\!\!-\!\!0.3407i & \ldots  \\                
   \ldots&     \ldots              & \ldots                  & \ldots          
\end{array}\right], \nonumber
\label{R23}
\end{eqnarray}

\begin{eqnarray}
\!\!\!\!\!\!\!\!\!\!\!\!\!\!\!\!R_{2,4}^{\dagger}\!\!=\!\!\left[ \begin{array}{ccccc}
        1&    0                  &  0        &       0               &  \ldots        \\           
        0&  0.6865               &  0        &-0.5482\!\!+\!\!0.4777i&  \ldots        \\            
        0&    0                  & 1         &       0               &  \ldots        \\                 
        0&-0.5482\!\!-\!\!0.4777i&  0        & -0.6865               &  \ldots        \\             
  \ldots &  \ldots               &   \ldots  &      \ldots           &  \ldots         
\end{array}\right], \nonumber
\label{R24}
\end{eqnarray}

\begin{eqnarray}
\!\!\!\!\!\!\!\!R_{2,5}^{\dagger}\!\!=\!\!\left[ \begin{array}{ccccc}
   1   &         0             &  0     &   0                   &       \ldots    \\         
   0   &      0.8929           &  0     &-0.4239\!\!-\!\!0.1518i &       \ldots      \\       
\vdots &      \vdots           &  \ddots&      \vdots           &       \ldots       \\         
   0   &-0.4239\!\!+\!\!0.1518i &  0     &   -0.8929             &       \ldots        \\    
\ldots &        \ldots         &  \ldots&   \ldots              &       \ldots        \\    
\end{array}\right], \nonumber
\label{R25}
\end{eqnarray}

\begin{eqnarray}
\!\!\!\!\!\!\!\!R_{2,6}^{\dagger}\!\!=\!\!\left[ \begin{array}{ccccc}
1      &      0                 & \ldots    &             0             &        0 \\         
0      &    0.9594              &  \ldots   &   -0.1233\!\!+\!\!0.2536i &        0 \\         
\vdots &  \vdots                &  \ddots   &           \vdots          &     \vdots  \\  
0      &-0.1233\!\!-\!\!0.2536i &  \ldots   &          -0.9594          &         0 \\         
0      &      0                 &  \ldots   &              0            &         1 
\end{array}\right], \nonumber
\label{R26}
\end{eqnarray}

\begin{eqnarray}
\!\!\!\!\!\!\!\!R_{3,4}^{\dagger}\!\!=\!\!\left[ \begin{array}{ccccc} 
\ldots &       \ldots              &    \ldots               &    \ldots    \\                    
 \ldots&    -0.6776\!\!-\!\!0.4981i&  -0.5333\!\!+\!\!0.0914i&    \ldots    \\             
 \ldots&    -0.5333\!\!-\!\!0.0914i&   0.6776\!\!-\!\!0.4981i&    \ldots    \\             
\ldots &                \ldots     &     \ldots              &   \ldots     \\                     
\end{array}\right], \nonumber
\label{R34}
\end{eqnarray}

\begin{eqnarray}
\!\!\!\!\!\!\!\!R_{3,5}^{\dagger}\!\!=\!\!\left[ \begin{array}{ccccc}                       
\ldots     &          \ldots             &    \ldots & \ldots                   & \ldots      \\           
\ldots     &      0.4375                 &     0     &-0.3023\!\!+\!\!0.8469i    & \ldots  \\                   
\ldots     &           0                 &     1     & 0                        & \ldots   \\                  
\ldots     &      -0.3023\!\!-\!\!0.8469i &     0     &          -0.4375         & \ldots   \\                
 \ldots    &          \ldots             &    \ldots &          \ldots          & \ldots 
\end{array}\right], \nonumber
\label{R35}
\end{eqnarray}

\begin{eqnarray}
\!\!\!\!\!\!\!\!R_{3,6}^{\dagger}\!\!=\!\!\left[ \begin{array}{cclcc}          
    \ldots    &   \ldots                &    \ldots &   \ldots                    &  \ldots  \\          
     \ldots   &  0.8262                 &           & 0.0453\!\!-\!\!0.5615i     &  \ldots      \\           
     \ldots   &  \vdots                 &   \ddots  &   \vdots                    &  \vdots      \\          
     \ldots   & 0.0453\!\!+\!\!0.5615i &           &   -0.8262                   &   \ldots  \\            
    \ldots    &      \ldots             &    \ldots &      \ldots                 &   \ldots 
\end{array}\right], \nonumber
\label{R36}
\end{eqnarray}

\begin{eqnarray}
\!\!\!\!\!\!\!\!R_{3,7}^{\dagger}\!\!=\!\!\left[ \begin{array}{ccccc}
 \ldots &              \ldots          & \ldots     &                   \ldots  & \ldots    \\ 
 \ldots &                0             & \ldots     &                     0     &  0         \\
 \ldots &            0.9727            & \ldots     &                  -0.2320  &  0       \\
 \ldots &              \vdots          &\ddots      &                \vdots     &  \vdots   \\
 \ldots &             -0.2320          & \ldots     &                   -0.9727 &  0       \\
 \ldots &                0             & \ldots     &                     0     &  1       
\end{array}\right], \nonumber
\label{R37}
\end{eqnarray}

\begin{eqnarray}
\!\!\!\!\!\!\!\!\!\!\!\!\!\!R_{3,8}^{\dagger}\!\!=\!\!\left[ \begin{array}{ccccc}
 \ldots &              \ldots          & \ldots     &                   \ldots  & \ldots    \\ 
 \ldots &                0             & \ldots     &                     0     &  0         \\
 \ldots &            0.7485            & \ldots     &                     0     &   0.6631       \\
 \ldots &              \vdots          &\ddots      &                \vdots     &  \vdots   \\
 \ldots &               0              & \ldots     &                   1       &  0       \\
 \ldots &                0.6631        & \ldots     &                     0     &  -0.7485       
\end{array}\right], \nonumber
\label{R38}
\end{eqnarray}

\begin{eqnarray}
\!\!\!\!\!\!\!\!\!\!\!\!\!\!R_{4,5}^{\dagger}\!\!=\!\!\left[ \begin{array}{cccc}
 \ldots      &   \ldots               &      \ldots              & \ldots     \\       
  \ldots     & 0                      &   -0.6608\!\!+\!\!0.7506i &  \ldots      \\      
 \ldots      & -0.6608\!\!-\!\!0.7506i &  0                       &  \ldots       \\      
 \ldots      &\ldots                  &  \ldots                  &  \ldots              
\end{array}\right], \nonumber
\label{R45}
\end{eqnarray}

\begin{eqnarray}
\!\!\!\!\!\!\!\!\!\!\!\!\!\!R_{4,6}^{\dagger}\!\!=\!\!\left[ \begin{array}{ccccc}
         \ldots  &            \ldots           &   \ldots   &   \ldots              &    \ldots   \\       
         \ldots  &             0.0240          &      0     &-0.9996\!\!-\!\!0.0164i &    0   \\       
         \ldots  &                0            &      1     &             0         &    0   \\       
         \ldots  &     -0.9996\!\!+\!\!0.0164i &      0     &           -0.0240     &    0   \\       
         \ldots  &                0            &      0     &             0         &    1          
\end{array}\right], \nonumber
\label{R46}
\end{eqnarray}

\begin{eqnarray}
\!\!\!\!\!\!R_{4,7}^{\dagger}\!\!=\!\!\left[ \begin{array}{ccccc}
 \ldots  &            \ldots          & \ldots  &     \ldots   & \ldots          \\ 
 \ldots  &           0.8438           &  \ldots &      0.5367  & 0          \\
 \ldots  &            \ldots          &  \ddots &       \ldots & \ldots        \\ 
 \ldots  &          0.5367            & \ldots  &      -0.8438 & 0                       \\
 \ldots  &              0             & \ldots  &          0   & 1        
\end{array}\right], \nonumber
\label{R47}
\end{eqnarray}

\begin{eqnarray}
\!\!\!\!\!\!R_{4,8}^{\dagger}\!\!=\!\!\left[ \begin{array}{ccccc}
 \ldots  &            \ldots          & \ldots  &     \ldots      & \ldots        \\ 
 \ldots  &            0.9255          &  \ldots &        0        &  0.3788       \\
 \ldots  &            \ldots          &  \ddots &       \ldots    & \ldots        \\ 
 \ldots  &              0             & \ldots  &        1        &    0          \\
 \ldots  &             0.3788         & \ldots  &        0        & -0.9255        
\end{array}\right], \nonumber
\label{R48}
\end{eqnarray}

\begin{eqnarray}
\!\!\!\!\!\!R_{5,6}^{\dagger}\!\!=\!\!\left[ \begin{array}{cccc}
 \ldots  &         \ldots                & \ldots                  &     \ldots        \\ 
 \ldots  &             0.0               & -0.6606\!\!-\!\!0.7507i &      0            \\ 
 \ldots  &       -0.6606\!\!+\!\!0.7507i &       0.0               &      0            \\
 \ldots  &            0                  &      0                  &      1         
\end{array}\right], \nonumber
\label{R56}
\end{eqnarray}

\begin{eqnarray}
\!\!\!\!\!\!R_{5,7}^{\dagger}\!\!=\!\!\left[ \begin{array}{ccccc}
 \ldots  &              \ldots                     & \ldots   &              &  \ldots        \\ 
 \ldots  &                   0.8615                & \ldots   &  0.5077      &   0      \\ 
 \ldots  &                    \ldots               &  \ddots  &     \ldots   &   \ldots   \\
 \ldots  &                0.5077                   &  \ldots  &    -0.8615   &   0        \\
  \ldots &              0                          & \ldots   &      0       &   1        
\end{array}\right], \nonumber
\label{R57}
\end{eqnarray}
         
\begin{eqnarray}
\!\!\!\!\!\!R_{5,8}^{\dagger}\!\!=\!\!\left[ \begin{array}{ccccc}
 \ldots  &              \ldots        & \ldots   &              &  \ldots        \\ 
 \ldots  &               0.2894       & \ldots   &      0       &   -0.9572      \\ 
 \ldots  &              \ldots        &  \ddots  &   \ldots     &   \ldots   \\
 \ldots  &                0           &  \ldots  &      1       &   0        \\
  \ldots &              -0.9572       & \ldots   &      0       &   -0.2894        
\end{array}\right], \nonumber
\label{R58}
\end{eqnarray}

\begin{eqnarray}
\!\!\!\!\!\!\!\!\!\!\!\!R_{6,7}^{\dagger}\!\!=\!\!\left[ \begin{array}{ccccc}
 \ldots  &  \ldots               & \ldots                 &     \ldots              &  \ldots   \\ 
 \ldots  &     1                 &      0                 &      0                  &  0 \\ 
 \ldots  &     0                 &      0.0               &      1.00               &  0 \\
 \ldots  &     0                 &      1.00              &      0.0                &  0  \\
 \ldots  &     0                 &      0                 &      0                  &  1
\end{array}\right]. \nonumber
\label{R67}
\end{eqnarray}

\begin{eqnarray}
\!\!\!\!\!\!\!\!\!\!\!\!R_{6,8}^{\dagger}\!\!=\!\!\left[ \begin{array}{ccccc}
 \ldots  &  \ldots               & \ldots                 &     \ldots              &  \ldots   \\ 
 \ldots  &     1                 &      0                 &      0                  &  0 \\ 
 \ldots  &     0                 &      0.7071            &      0                  &  -0.7071 \\
 \ldots  &     0                 &      0                 &      1                  &  0  \\
 \ldots  &     0                 &     -0.7071            &      0                  &  -0.7071
\end{array}\right]. \nonumber
\label{R68}
\end{eqnarray}

\begin{eqnarray}
\!\!\!\!\!\!\!\!\!\!\!\!R_{7,8}^{\dagger}\!\!=\!\!\left[ \begin{array}{ccccc}
 \ldots  &  \ldots               & \ldots                 &     \ldots              &  \ldots   \\ 
 \ldots  &     1                 &      0                 &      0                  &  0 \\ 
 \ldots  &     0                 &      1                 &      0                  &  0 \\
 \ldots  &     0                 &      0                 &      0.0                &  0.5921\!\!-\!\!0.8058i  \\
 \ldots  &     0                 &      0                 &      1.00               &  0.0
\end{array}\right]. \nonumber
\label{R78}
\end{eqnarray}

Specifying the parameters ($\alpha,\beta,\gamma$, $\delta$) in each step of the decomposition we conclude the the unambiguous discrimination protocol.  The parameter's values are given in Table I.

\begin{table}
\caption{\label{tab:table2}A specification of the parameters $\alpha,\beta,\gamma$, and $\delta$  involved in the sequence of the two-level operations in the example. The angles $\alpha,\beta,\gamma$, and $\delta$ are given in degrees.}
\begin{ruledtabular}
\begin{tabular}{rrrrr}
          &$\alpha$ & $\beta$   & $\gamma/2$& $\delta$   \\
\hline
$R_{1,2}$  &   90.0   &      0.0  &     45.0  & 180.0        \\
$R_{1,3}$  &   90.0   &      0.0  &    35.27  & 180.0        \\
$R_{1,4}$  &   90.0   &      0.0  &     30.0  & 180.0        \\
$R_{2,3}$  &  -90.0   & -2.2066   &  -36.9938 & -128.4034        \\
$R_{2,4}$  &  -90.0   &-138.9380  &   46.6463 & -41.0619        \\
$R_{2,5}$  &   90.0   & 160.2880  &   26.76   & 19.7119        \\
$R_{2,6}$  &  -90.0   &-115.9253  & 16.3825   & -64.0746     \\
$R_{3,4}$  &  -90.0   & -26.6549  & 32.7541   & 134.0108     \\
$R_{3,5}$  &  -90.0   & -109.6455 & 64.0555   & -70.3544      \\
$R_{3,6}$  &  -90.0   & -94.6115  & -34.2896  & -85.3884    \\
$R_{3,7}$  &   90.0   & 180.00    & 13.4187   & 0.0     \\
$R_{3,8}$  &   90.0   &  0.0      & 41.5393   & 180.00   \\
$R_{4,5}$  &  -90.0   & -131.3609 & 90.00     & -48.6390   \\
$R_{4,6}$  &   90.0   & 179.1897  & 88.6247   & 0.8102    \\
$R_{4,7}$  &   90.0   & 0.0       & 32.4564   & 180.0      \\
$R_{4,8}$  &   90.0   & 0.0       & 22.2561   & 180.0      \\
$R_{5,6}$  &   90.0   & 131.3456  & 90.0      & 48.6543    \\
$R_{5,7}$  &   90.0   & 0.0       & 30.5145   & 180.0      \\
$R_{5,8}$  &   90.0   & 180.0     & 73.1779   & 0.0   \\
$R_{6,7}$  &   90.0   & 0.0       & 90.0      & 180.0      \\
$R_{6,8}$  &   90.0   & 180.0     & 45.0      & 0.0    \\
$R_{7,8}$  &26.8469   & 0.0       & 90.0      & 53.6938    \\
\end{tabular}
\end{ruledtabular}
\end{table}

\section{Conclusions}

We showed a general algorithm to perform the optimal discrimination of
$N$ linearly independent non-orthogonal pure quantum states by means of semidefinite programming and norm minimization. 
In addition, we presented a simple computational code that takes as input the ensemble of non-orthogonal states and outputs the best set of discriminable states and the sequence of two-level
rotation matrices. As a numerical example, we studied
an USD attack to the BB84 protocol.

\section*{Acknowledgments}
We are grateful to Marcello A. Talarico, Fernando G. S. L. Brand\~ao, S. P\'adua and C. Saavedra for  valuables discussions. This work was supported by FAPEMIG, CNPq, Instituto do Mil\^enio de Informa\c c\~ao Qu\^antica.

\appendix

\section{Determination of the inconclusive probability amplitudes}

In this appendix we supply the procedure to calculate the inconclusive probability amplitudes in the final configuration, i.e,
$g_i$, for   $i=N+1$ to $\frac{1}{2}N(N+3)-1$.
We detail the procedure for the case $N=3$, and then discuss the
generalization, presenting a MATLAB code for arbitrary $N$.

Let's consider three non-orthogonal quantum states in a ladder form. It is convenient to factorize a global phase and  set the coefficient multiplying the the ket $\ket{1}$ as real,
 for all states,  then:
\begin{eqnarray}
\left|Q_1\right\rangle&=&\phantom{cc}\left|1\right\rangle \nonumber \\
\left|Q_2\right\rangle&=&a_1\left|1\right\rangle+a_2e^{i\theta_1}\left|2\right\rangle 
\nonumber \\
\left|Q_3\right\rangle&=&
a_3\left|1\right\rangle+a_4e^{i\theta_2}\left|2\right\rangle+a_5e^{i\theta_3}\left|3\right\rangle ~~,
\label{3-initial}
\end{eqnarray}
The  final  discriminable  configuration in the extended Hilbert space is
\begin{eqnarray}
\left|Q_{1f}\right\rangle&=&g_1\left|1\right\rangle+g_4\left|4\right\rangle+g_5\left|5\right\rangle 
\nonumber \\
\left|Q_{2f}\right\rangle&=&g_2\left|2\right\rangle+g_6\left|4\right\rangle+g_7\left|5\right\rangle 
\nonumber \\
\left|Q_{3f}\right\rangle&=&
g_3\left|3\right\rangle+g_8\left|4\right\rangle. 
\label{3-finais}
\end{eqnarray}
The conclusive probability amplitudes $g_1,g_2,g_3$ are determined via SDP. The inconclusive probability amplitudes $g_4$ to $g_8$ must be determined preserving  normalization and scalar products  among the vectors. From normalization we have,
\begin{eqnarray}
\label{norma-g5}
|g_1|^2+|g_4|^2+|g_5|^2=1  \\
\label{norma-g7}
|g_2|^2+|g_6|^2+|g_7|^2=1   \\
\label{norma-g8} 
|g_3|^2+|g_8|^2=1 ~~,    
\end{eqnarray}
and the scalar products  are
\begin{eqnarray}
\left\langle Q_{1f}|Q_{2f}\right\rangle&=&\left\langle 
Q_{1}|Q_{2}\right\rangle, \nonumber \\
\left\langle Q_{1f}|Q_{3f}\right\rangle&=&\left\langle 
Q_{1}|Q_{3}\right\rangle, \nonumber\\
\left\langle Q_{2f}|Q_{3f}\right\rangle&=&\left\langle 
Q_{2}|Q_{3}\right\rangle.
\label{escalar-3}
\end{eqnarray}
 (\ref{escalar-3}) can be written in the form
\begin{eqnarray}
\label{Q1_Q2}
g_4^*g_6+g_5^*g_7&=&a_1 \\
\label{Q1_Q3}
g_4^*g_8 &=&a_3  \\
\label{Q2_Q3}
g_6^*g_8 &=&a_2a_4e^{i(\theta_2-\theta_1)}+a_1a_3 ~~.
\end{eqnarray}
From (\ref{Q1_Q3}) it is observed that $g_4$ and $g_8$ can be taken as  real. Due to (\ref{norma-g8}) we have $g_8=+\sqrt{1-|g_3|^2}$. Substituting $g_8$ in (\ref{Q1_Q3}) we get $g_4=+a_3/\sqrt{1-|g_3|^2}$. Substituting $g_8$ in (\ref{Q2_Q3}) we have
\begin{eqnarray}
g_6=\frac{[a_2a_4\cos\Theta+a_1a_3]-i[a_2a_4\sin\Theta]}{\sqrt{1-|g_3|^2}}~~.
\end{eqnarray}
with $\Theta=\theta_2-\theta_1$.

$g_5$ and $g_7$ are still undetermined. Writing them as $g_5=g_5^R+ig_5^{Im}$ e $g_7=g_7^R+ig_7^{Im}$, and using (\ref{norma-g5}), (\ref{norma-g7}), (\ref{Q1_Q2}), then
\begin{eqnarray}
\label{g7-real}
g_7^R&=&\pm\sqrt{b-(g_7^{Im})^2},\\
\label{g5-real}
g_5^R&=&\pm\sqrt{a-(g_5^{Im})^2},\\
\label{sistema-g5}
g_7^Rg_5^R+g_5^{Im}g_7^{Im} &=&c, \\
\label{sistema-g7}
g_5^Rg_7^{Im}-g_5^{Im}g_7^R &=&d,
\end{eqnarray}
with  $a, b, c$ and $d$ reals  given by
\begin{eqnarray}
a&=&1-|g_1|^2-|g_4|^2 , \nonumber  \\
b&=&1-|g_2|^2-|g_6|^2 ~~, \nonumber \\
c&=&a_1-\frac{a_3[a_1a_3+a_2a_4\cos\Theta]}{1-|g_3|^2} ~~, \nonumber \\
d&=&\frac{a_3a_2a_4\sin\Theta}{1-|g_3|^2} ~~.
\end{eqnarray} 
Substituting (\ref{g7-real}) and (\ref{g5-real}) into (\ref{sistema-g5}) and (\ref{sistema-g7}) we have
\begin{eqnarray}
\label{sistema-1}
\!\!\!\!\!\!\!\!\!\!\!\!\!\!\!\!\!\!\!\sqrt{a-(g_5^{Im})^2}~\sqrt{b-(g_7^{Im})^2}+g_5^{Im}g_7^{Im}  &=& c~~, \\
\label{sistema-2}
\!\!\!\!\!\!\!\!\!\!\!\!\!\!\!\!\!\!\!\sqrt{a-(g_5^{Im})^2}~g_7^{Im}-g_5^{Im}~\sqrt{b-(g_7^{Im})^2} &=& d~~, 	
\end{eqnarray}
From (\ref{sistema-1}) we have
\begin{eqnarray}	
\!\!\!\!\!\!\!\!\!g_7^{Im}\!\!\!=\!\frac{2cg_5^{Im}\!\pm\!\sqrt{(2cg_5^{Im})^2\!-\!4(b(g_5^{Im})^2+c^2\!-\!ab)a}}{2a},
\label{g7m-valendo}	
\end{eqnarray}
and from (\ref{sistema-2})
\begin{eqnarray}	
(g_7^{Im})^4&+&\left[\frac{(4d^2-2ab)(g_5^{Im})^2+2ad^2}{a}\right]~(g_7^{Im})^2+  \nonumber \\
	          &+& \left[\frac{b^2(g_5^{Im})^4-6d^2b(g_5^{Im})^2+d^4}{a}\right]\!\!=\!0~.
\label{g7m-polinomio}	
\end{eqnarray}

Substituting (\ref{g7m-valendo}) into (\ref{g7m-polinomio}) we find that $g_5^{Im}$ must obey the following polynomial
\begin{eqnarray}
\!\!\!\!\!\!\!\!A(g_5^{Im})^8\!+\!B(g_5^{Im})^6\!+\!C(g_5^{Im})^4\!+\!D(g_5^{Im})^2\!+\!E\!\!=\!\!0~,
\label{g5m-polinomio}
\end{eqnarray}
with real  coefficients
\begin{eqnarray}
&&\mathbf{A}=\left[(4d^2-4ab)(c^2-ab)-4c^2d^2\right]^2~~,  \nonumber \\ 
&&\mathbf{B}=\left[(4d^2-4ab)(c^2-ab)+4c^2d^2\right]\times  \nonumber \\
&&\times\left[8a^2b(c^2-ab-d^2)\right]-64c^2d^2(c^2-ab)(2a^2b)~~,\nonumber\\
&&\mathbf{C}=\left[a^2d^4+a^2(c^2-ab)^2-2d^2a^2(c^2-ab)\right]\times \nonumber \\ 
&&\times \left[(4d^2-4ab)(c^2-ab)+4c^2d^2\right] + \nonumber \\
&&+\left[4a^2b(c^2-ba-d^2)\right]^2+ \nonumber \\
&&+64c^2d^2(c^2-ab)(d^2+ab)a^2~~,  \nonumber \\
&&\mathbf{D}=\left[a^2d^4+a^2(c^2-ab)^2-2d^2a^2(c^2-ab)\right]\times 
\nonumber \\
&&\times \left[4a^2b(c^2-ba-d^2)\right]~~, \nonumber\\
\label{coef}
&&\mathbf{E}=\left[a^2d^4+a^2(c^2-ab)^2-2d^2a^2(c^2-ab)\right]^2.
\end{eqnarray}
The roots of this  polynomial are easily obtained \cite{Horn}.
We choose for $g_5^{Im}$ any real root, such that $0\leq (g_5^{Im})^2\leq a$. Once $g_5^{Im}$ is determined, we calculate $g_7^{Im}$ from (\ref{g7m-valendo}). Then (\ref{g7-real}) and (\ref{g5-real}) are used to calculate $g_7^R$ and $g_5^R$. Now all the parameters in the final configuration are determined.

It is straightforward to extend this procedure for arbitrary $N$.
Summarizing: we determine the conclusive amplitudes, $g_1$ to $g_N$, 
by SDP; then, starting from the last inconclusive amplitude,
$g_{[\frac{1}{2}N(N+3)-1]}$, we use the relations for normalization
and scalar products and thus determine all the remaining parameters,
except for $g_{2N-1}$ and $g_{3N-2}$, which are obtained by means of
the polynomial
\begin{eqnarray}
\!\!\!\!\!\!\!A(g_{2N\!-\!1}^{Im})^8\!&+&\!B(g_{2N\!-\!1}^{Im})^6\!+\!C(g_{2N\!-\!1}^{Im})^4\!+ \nonumber\\
                                        &+&\!D(g_{2N\!-\!1}^{Im})^2\!+\!E=0.
\label{polinomio-geral}
\end{eqnarray}

\begin{verbatim}

% Unambiguos State Discrimination
% Wilson R.M. Rabelo (wilson@unifap.br)
% 09/2005 UFMG  Quantum Information Group
%%%%%%%%%%%%%%%%%%%%%%%%%%%%%%%%%%%%%%%%%
%             BEGIN INPUT - BB84
%%%%%%%%%%%%%%%%%%%%%%%%%%%%%%%%%%%%%%%%%
%Number of states
N=4;
%Hilbert space dimension
% (at least 2N-1)
dim=8;
%Probability of each state in the
%                       ensemble
Mi=-1*[0.25; 0.25; 0.25; 0.25];
zero=[1;0];
one=[0;1];
lr=(zero+one)/sqrt(2);
ll=(zero-one)/sqrt(2);
cr=(zero+i*one)/sqrt(2);
cl=(zero-i*one)/sqrt(2);
lr2=kron(lr,lr);
ll2=kron(ll,ll);
cr2=kron(cr,cr);
cl2=kron(cl,cl);
lr3=kron(lr2,lr);
ll3=kron(ll2,ll);
cr3=kron(cr2,cr);
cl3=kron(cl2,cl);
%Input sates QII= Q_1,...,Q_N
QII=[lr3 ll3 cr3 cl3];

%%%%%%%%%%%%%%%%%%%%%%%%%%%%%%%%%%%%%%%%
%            END INPUT
%%%%%%%%%%%%%%%%%%%%%%%%%%%%%%%%%%%%%%%%
 % BEGINNING:
 %%%%%%%%%%%%%%%%%%%%%%%%%%%%%%%%%%%%%%%%
 %N initial states:
 disp('N initial states:')
 disp(QII)
 %Scalar product for N initial states:
 
 for jL=1:N
   for iL=1:N
    Esc_QII=QII(1:dim,iL)'*QII(1:dim,jL);
    esc_QII(iL,jL)=Esc_QII;
   end
 end
%%%%%%%%%%%%%%%%%%%%%%%%%%%%%%%%%%%%%%%%%%
%     Rewrite the entry states 
%      in  ladder form (Uo)            
%%%%%%%%%%%%%%%%%%%%%%%%%%%%%%%%%%%%%%%%%%
%for ig=1:2*N-1
for ig=1:dim
  for jg=1:N
      c(ig,jg)=0.0;
  end
end
c(1,1)=1.0;
for ittt=1:N-1
   for j=ittt+1:N
      Sfat_w=0.0;
      if ittt > 1
        for xxw=1:ittt-1
 Sfat_w=Sfat_w+conj(c(ittt-xxw,ittt))*c(ittt-xxw,j);
        end
      end
      if N==2
        c(ittt,j)=esc_QII(ittt,j);
 c(ittt+1,ittt+1)=sqrt(1-conj(c(ittt,ittt+1))*c(ittt,ittt+1));
        continue
      end
      if N >= 3
          if ittt==1
            c(ittt,j)=esc_QII(ittt,j);
          else
       c(ittt,j)=(esc_QII(ittt,j)-Sfat_w)/c(ittt,ittt);
          end
      end
  end
  Swfat=0.0;
  for xo=1:ittt
      Swfat=Swfat+conj(c(xo,ittt+1))*c(xo,ittt+1);
  end
  c(ittt+1,ittt+1)=sqrt(1-Swfat);
end
              
Q=c;
format short
AAA=zeros(dim);
for ux=1:N
AAA=AAA +Q(1:dim,ux)*QII(1:dim,ux)';
end
[V_0,Sigma_0,W_0]=svd(AAA);
disp('The unitary matrix Uo ')
disp('to put states in ladder form :')
Uo=V_0*W_0'
disp('Test  Uo, i.e, (Uo*)(Uo)=')       
disp(Uo*Uo')
disp('Initial configuration:[Q1...QN]')
disp(' in  ladder form:')
for iu=1:N
Qcc(:,iu)=Uo*QII(1:dim,iu);
end
Q
pause
format long
%%%%%%%%%%%%%%%%%%%%%%%%%%%%%%%%%%%%%%%%%
%Scalar product for N states 
%    in  ladder form:
 for jL=2:N
   for iL=1:jL-1
      Esc_Q=Q(1:N,iL)'*Q(1:N,jL);
      esc_Q(iL,jL)=Esc_Q;
   end
 end
 %%%%%%%%%%%%%%SDP Approach%%%%%%%%%%%%%%
  c=sdpvar(N,1);
  p=sdpvar(N,1);
  F_0=eye(dim);
  Qt=Q*inv(Q'*Q);
  D=F_0;
  for ib=1:N 
    D=D-p(ib)*Qt(1:dim,ib)*Qt(1:dim,ib)';
  end
  yalmip('info');
  F= set(D > 0);
  for ikp=1:N
    F=F+set(p(ikp)>0);
  end
  solvesdp(F,Mi'*p);
  P=double(p)
 %%%%%%%%%%%%%%%%%%%%%%%%%%%%%%%%%%%%%%%
  for jj=1:N
      g(jj)=sqrt(P(jj));
  end
  %%%%%%%%%%%%%%%%%%%%%%%%%%%%%%%%%%%%%%%
  disp('Conclusive amplitudes via SDP :')
  g
  %%%%%%%%%%%%%%%%%%%%%%%%%%%%%%%%%%%%%%%
  pause

  for ii=N+1:((N/2)*(N+3))-1
      g(ii)=0.0;
  end    
  
  g(((N/2)*(N+3))-1)=sqrt(1-g(N)^2);
  
 x=1; Sfatg=0;jjj=0;
 w=3;
 for jL=N:-1:3
  j=((N/2)*(N+3))-x; 
  jjj=jjj+1;
  k=jjj;
  for kL=jL-1:-1:1
    if kL > 1
      k=k+1;
      j=j-k;
      if w > 3
       Sfatg=0;
       P=0;
       fatg=0;
     for f=((w/2)*(w-5))+5:((w/2)*(w-3))+1
         P=P+1;
         E(f)=g(((N/2)*(N+3))-f)/g(((N/2)*...
                   (N+3))-x);
         fatg=conj(g(j-P))*E(f);
         Sfatg=Sfatg+fatg;
       end
      end
      g(j)=((esc_Q(kL,jL))/(g(((N/2)*...
                   (N+3))-x)))-Sfatg;
      g(j)=conj(g(j));
      else
        j=j-k;
         if w > 3
          Sfatg=0;
          P=0;
          fatg=0;
        for f=((w/2)*(w-5))+5:((w/2)*(w-3))+1
            P=P+1;
            E(f)=g(((N/2)*(N+3))-f)/g(((N/2)*...
                   (N+3))-x);
            fatg=conj(g(j-P))*E(f);
            Sfatg=Sfatg+fatg;
         end
          end
        g(j)=((esc_Q(kL,jL))/(g(((N/2)*...
                   (N+3))-x)))-Sfatg;
        g(j)=conj(g(j));
       end
     end
      if N==3 
             continue
      end
      if jL > 3
         w=w+1;
         fat=0;
         Sfat1=1-g(jL-1)*conj(g(jL-1));
         x=((w/2)*(w-5))+4;
       for t=((w/2)*(w-5))+5:((w/2)*(w-3))+1
         fat=fat+conj(g(((N/2)*(N+3))-t))*...
                   g(((N/2)*(N+3))-t);
       end
          g(((N/2)*(N+3))-x)=sqrt(Sfat1-fat);
      else
            continue
      end
  end
  fatp=0.0;
  for iik=N+1:2*N-2
      fatp=fatp+g(iik)*conj(g(iik));
  end
  a=1-g(1)^2-fatp;
  
  fatpp=0.0;
  for iiik=2*N:3*N-3
      fatpp=fatpp+g(iiik)*conj(g(iiik));
  end
  b=1-g(2)^2-fatpp;
  
  fatig=0.0;
  for pp=2*N:3*N-3
      fatig=fatig+conj(g(pp+(1-N)))*g(pp);
  end
  g2N_1_g3N_2=esc_Q(1,2)-fatig;
  c=real(g2N_1_g3N_2);
  d=imag(g2N_1_g3N_2);
  
if d==0.0   
      g(2*N-1)=sqrt(a);
      g(3*N-2)=sqrt(b);
      format short;
      
elseif( (c < 1.0e-010) & (c >-0.1e-05))
      g(3*N-2)=sqrt(b);
      g_2N_1_im=-sqrt(a);
      g(2*N-1)=complex(0,g_2N_1_im);
      format short;
     
elseif ((d < 1.0e-010) & (d > -0.1e-05))  
      g(2*N-1)=sqrt(a);
      g(3*N-2)=sqrt(b);
      format short;
      
else
 %%%%% Polynomial%%%% 
  AA=[(4*d^2-4*a*b)*(c^2-a*b)-4*c^2*d^2]^2;

  BB=[(4*d^2-4*a*b)*(c^2-a*b)+4*c^2*d^2]*...
  [8*a^2*b*(c^2-a*b-d^2)]-64*c^2*d^2*...
  (c^2-a*b)*(2*a^2*b);

  CC=[a^2*d^4+a^2*[(c^2-a*b)^2]-2*d^2*a^2*...
  (c^2-a*b)]*[(4*d^2-4*a*b)*(c^2-a*b)+...
  4*c^2*d^2]+[4*a^2*b*(c^2-a*b-d^2)]^2+...
  64*c^2*d^2*(c^2-a*b)*(d^2+a*b)*a^2;

  DD=[a^2*d^4+a^2*[(c^2-a*b)^2]-2*a^2*...
  d^2*(c^2-a*b)]*4*a^2*b*(c^2-a*b-d^2);

  EE=[a^2*d^4+a^2*(c^2-a*b)^2-2*...
  a^2*d^2*(c^2-a*b)]^2;
 %%%%%%%%%%%%%%%%%%%%%%%%%%%%%%%%%%%%%%%%
  Pii=[AA 0 BB 0 CC 0 DD 0 EE];
    rr=roots(Pii);
    
    for ih=1:8
      if isreal(rr(ih))
       if (rr(ih) < 1.0) & (rr(ih) > 0.0)
         g_2N_1_im=rr(ih);
         g3N_2_im=[2*c*g_2N_1_im+...
          sqrt((2*c*g_2N_1_im)^2-4*a*...
           [b*g_2N_1_im^2+c^2-a*b])]/(2*a);
         g_2N_1_R=-sqrt(a-g_2N_1_im^2);
         g3N_2_R=-sqrt(b-g3N_2_im^2);
            %%%%%%%%%%%%%%%%%%%%%%%%%%%%%
        ccc=g3N_2_R*g_2N_1_R+...
          g3N_2_im*g_2N_1_im;
        ddd=g_2N_1_R*g3N_2_im-...
          g_2N_1_im*g3N_2_R;
        ai=g_2N_1_R;
        bi=g_2N_1_im;
        g(2*N-1)=complex(ai,bi);  
        aii=g3N_2_R;
        bii=g3N_2_im;
        g(3*N-2)=complex(aii,bii);
        format short;
          else
            continue
          end
      else
       if ih==8
         disp('roots=')
         disp(rr)
         disp('Polynomial roots ')
         disp(' are not real!')
         disp('Check input states!')
         disp('Input states cannot') 
         disp(' be linearly dependent.') 
         return
        else
            continue
        end
      end
    end  
end
disp('  ')
disp('Conclusive and inconclusive')
disp(' amplitude probabilities :')
  g
 pause
 
%%%%Building final vectors Qf %%%%%
if dim==2*N-1;
  xdim=2*N-1;
else  
  xdim=dim;
end

for uk=1:N;
    for ui=1:xdim;
        Qf(ui,uk)=0.0;
    end
end  
%-------
jj=0 ;   
for iiiv=1:N;

    jj=jj+1;
   for uk=1:N;
        if uk==jj;
            Qf(uk,iiiv)=g(uk);
        else uk<N+1 ;
            continue
        end
   end
end
S=N;
uk=0;
iyy=2*N-1;
for j=1:N;
    uk=uk+1;
        if j==1 ;
            for  zk=S+1:S+(N-j);
                Qf(zk,uk)=g(zk) ;   
            end
 
        elseif  j<= N-1  ;
         xx=uk-2;
          for zk=S+1:S+(N-j)+1;
           zmim=S+1;
           Qf(iyy-(N-uk)+(zk-zmim)-...
             xx,uk)=g(zk);             
          end
      else
            zk=((N/2)*(N+3))-1;
            Qf(N+1,uk)=g(zk);
      end
      S=zk;
end
      Qf
      pause
      
%%%%%%%%%%%External_product A'%%%%%%%%%%%
    A=zeros(xdim);
    for ux=1:N
        A=A +Qf(1:xdim,ux)*Q(1:xdim,ux)';
    end
    disp('A=')
    disp(A)
        [V,Sigma,W]=svd(A) ;
        disp('The unitary matrix U1 :')
        U1=V*W'
    disp('Test  U1, i.e, (U1*)(U1)=')       
    disp(U1*U1')
    disp('The final configuration:')
    disp('[Q1f Q2f Q3f ... QNf]')
        for iu=1:N
        Qfc(:,iu)=U1*Q(1:xdim,iu);
        end
        disp(Qfc)

  disp('decomposing  (U1)*(Uo)')
  disp('U=U1*Uo')
  disp(U1*Uo)
  pause

%%%%Decomposing U=(U1)(Uo)%%%%%%%

U_aux=U1*Uo;
k=0;
for ifg=1:xdim-2 ;
  for j=ifg+1:xdim ;
     k=k+1;
     a=U_aux(ifg,ifg);
     b=U_aux(j,ifg);
     c=sqrt(conj(a)*a+conj(b)*b);
     a=a/c;
     b=b/c;
    if ((conj(b)*b)<0.0001) & (j==ifg+1);
       V=eye(xdim);
       R(:,:,ifg,j)=V;   
       U_aux=V*U_aux;
       continue
    elseif ((conj(b)*b)<0.0001) & (j>ifg+1);
       V=eye(xdim);
       V(ifg,ifg)=conj(a);
       R(:,:,ifg,j)=V ;  
       U_aux=V*U_aux;
       continue
    else ((conj(b)*b)>0.00000001);
       V=eye(xdim);
       V(ifg,ifg)=conj(a);
       V(ifg,j)=conj(b);
       V(j,ifg)=b;
       V(j,j)=-a;
       R(:,:,ifg,j)=V ;       
       U_aux=V*U_aux;
        continue
     end
   end
 end
     V_aux=U_aux;
     ifg=xdim-1;
     j=ifg+1;
     k=k+1;
    V=eye(xdim);
    V(ifg,ifg)=conj(V_aux(ifg,ifg));
    V(ifg,j)=conj(V_aux(j,ifg));
    V(j,ifg)=conj(V_aux(ifg,j));
    V(j,j)=conj(V_aux(j,j));
    R(:,:,ifg,j)=V;
    for ifg=1:xdim-1 ;
        for j=ifg+1:xdim ;
            if j==ifg ;
                continue
            else
             Rotation(:,:,ifg,j)=R(:,:,ifg,j)';
             disp('Rotation'),disp([ifg j])
             disp(Rotation(:,:,ifg,j))
            end
        end
        pause
    end
disp('Test Rotations')
disp('R_(d-1,d)R_(d-2,d)...R(1,3)R_(1,2)U=')
    U_final=V*V_aux;
    disp(U_final)
%%%% END %%%%%%%


\end{verbatim}

\end{document}